# Ethyl alcohol and sugar in comet C/2014 Q2 (Lovejoy)

Nicolas Biver,[1*] Dominique Bockelée-Morvan,[1] Raphaël Moreno,[1] Jacques Crovisier,[1] Pierre Colom,[1] Dariusz C. Lis,[2,3] Aage Sandqvist,[4] Jérémie Boissier,[5] Didier Despois,[6] Stefanie N. Milam[7]

[1]LESIA, Observatoire de Paris, PSL Research University, CNRS, UPMC, Université Paris-Diderot, 5 place Jules Janssen, F-92195 Meudon, France.

[2]LERMA, Observatoire de Paris, PSL Research University, CNRS, Sorbonne Universités, UPMC Univ. Paris 06, 61 av. de l'Observatoire, F-75014, Paris, France.

[3]Cahill Center for Astronomy and Astrophysics 301-17, California Institute of Technology, Pasadena, CA 91125, USA.

[4]Stockholm Observatory, AlbaNova University Center, SE-106 91 Stockholm, Sweden.

[5]IRAM, 300, rue de la Piscine, F-38406 Saint Martin d'Hères, France.

[6]LAB, Univ. Bordeaux, CNRS, UMR 5804, 2 rue de l'Observatoire, F-33271 Floirac, France.

[7]NASA Goddard Space Flight Center, Astrochemistry Laboratory, Washington, USA.

[*]Corresponding author. E-mail: nicolas.biver@obspm.fr

## Abstract

The presence of numerous complex organic molecules (COMs; defined as those containing 6 or more atoms) around protostars shows that star formation is accompanied by an increase of molecular complexity. These COMs may be part of the material from which planetesimals and ultimately planets formed *(1)*. Comets sample some of the oldest and most primitive material in the solar system, including ices, and are thus our best window into the volatile composition of the solar proto-planetary disk. Molecules identified to be present in cometary ices include water, simple hydrocarbons, oxygen, sulphur and nitrogen-bearing species, as well as a few complex organic molecules, such as ethylene glycol and glycine *(2–4)*. Here, we report the detection of twenty-one molecules in comet C/2014 Q2 (Lovejoy), including the first identification of ethyl alcohol (ethanol, $C_2H_5OH$) and the simplest monosaccharide sugar glycolaldehyde ($CH_2OHCHO$) in a comet. The abundances of ethanol and glycolaldehyde, respectively 5 and 0.8% relative to methanol (0.12 and 0.02% relative to water), are somewhat higher than values measured in solar-type protostars. Overall, the high abundance of COMs in cometary ices supports formation through grain-surface reactions in the Solar System protoplanetary disk.

## *Introduction*

Comet C/2014 Q2 (Lovejoy) is a long period comet originating from the Oort Cloud, which passed its perihelion on 30 January 2015, at 1.290 astronomical units (AU) from the Sun. This comet was a naked-eye object in January and February 2015. At perihelion, its water production rate exceeded 20 metric tons/s. It was thus one of the most active comets in Earth orbit neighborhood since the passage in 1997 of the extraordinarily active comet C/1995 O1 (Hale-Bopp). Bright comets offer the opportunity to detect numerous species in their atmospheres and to search for new molecules outgassing from the nucleus ices.

Using the 30-m telescope of the Institut de Radioastronomie Millimétrique (IRAM), located on Pico Veleta in the Sierra Nevada (Spain), we conducted observations of the atmosphere of comet Lovejoy between 13–16 and 23–26 January 2015, when the comet was the brightest and the most productive, at a distance of 0.6 AU from the Earth, and of 1.3 AU from the Sun. Thanks to the versatility of the receivers and spectrometers at the IRAM 30-m telescopes *(5)*, we were able to survey most of the 210–272 GHz ($\lambda \sim 1$ mm) spectral domain, with high spectral resolution and sensitivity. The angular resolution is $\sim$ 9" to 12" at these frequencies, corresponding to ~3300 to 5400 km at the distance of the comet. The targeted spectral range covers many molecular rotational lines, and has been successfully used to identify complex organic molecules in comets *(6–8)* and in star-forming regions, for example, in the studies of Belloche *et al. (9)* and Caux *et al. (10)*.

## *Results*

In the spectral survey of comet Lovejoy, we detected lines of twenty-one molecules (Table 1), two of which, ethyl alcohol (ethanol, $C_2H_5OH$) and glycolaldehyde ($CH_2OHCHO$), were detected for the first time in a comet. We secured the detection of ethanol and glycolaldehyde by averaging the strongest lines present in the observed spectral domain. Ethanol is detected through lines of both *trans* and *gauche*-conformers. Most other detected COMs and simple organic molecules, such as ethylene glycol (($CH_2OH)_2$), methyl formate ($HCOOCH_3$), formamide ($NH_2CHO$), formic acid ($HCOOH$) and acetaldehyde ($CH_3CHO$), were first detected in comet Hale-Bopp, and later confirmed in other comets *(6–8)*. Figure 1 shows spectra of eight organic species detected in comet Lovejoy.

Relative abundances can be estimated by deriving molecular production rates from the observed line intensities. In this active comet, most molecules within the field of view are close to Local Thermal Equilibrium (LTE). To take into account departures from LTE in the less dense parts of the coma, due to radiative decay, the line intensities were analyzed using an excitation model *(11, 12)*. From the numerous detected methanol lines, we derive a rotational temperature of $68.0 \pm 0.7$ K and a gas temperature of 73 K. Assuming isotropic outgassing, and using the gas velocity of 0.8 km/s derived from the spectrally resolved strong lines, we determined the production rate for all observed molecules *(5)*. Table 1 provides abundances relative to water, using water production rates of $Q_{H2O} = 5 \times 10^{29}$ and $6 \times 10^{29}$ molecules s$^{-1}$, for the periods 13–16 and 23–26 January 2015, respectively. These values are deduced from observations of the OH radical and $H_2O$ performed with the Nançay radio telescope and Odin space observatory, respectively, in January 2015. The conformers of ethanol are found to be in relative proportion *trans/gauche* of $1.5 \pm 0.3$. Because of the low energy barrier between the two ethanol conformers ($\sim$ 57 K), trans to gauche inter-conversion is rapid in the collisional region of the comet (less than 1 ps at 73 K), whereas inter-conversion through radiative decay is expected to be slow *(13)*. The *trans/gauche* ratio is close to the equilibrium value of 1.1 expected at the temperature of the sampled gas (73 K).

## *Discussion*

Production rates relative to water measured in comet Lovejoy are lower by a factor 2 to 3 than those in comet Hale-Bopp, except for $CH_3OH$ and $HCOOCH_3$, which show similar abundances, and for cyanoacetylene ($HC_3N$), which is depleted in Lovejoy by a factor of 10 *(6)*. Another notable exception is CO, which is a factor > 10 less abundant in Lovejoy, but Hale-Bopp belongs to the small family of CO-rich comets *(14)*. The depletion remains when comparing with abundances measured in comets other than Hale-Bopp (Table S2). Composition diversity is often observed in the comet population, but this is the first time that depletion is observed for such a large number of molecules in a comet.

Ethanol, glycolaldehyde, as well as the other COMs observed in comets, are found in warm ($\geq$ 100 K) regions surrounding newly born stars, at the time when the collapsing envelope feeding the star has not yet dissipated *(1)*. These regions, called Hot Corinos/Cores for low/high mass protostars, exhibit a rich molecular inventory dominated by saturated species. COMs may have been synthesized in the cold pre-stellar core period by grain-surface reactions, when all heavy-element bearing molecules were frozen onto grains *(15, 16)*. The efficiency of surface chemistry is increased during the warm-up phase, thanks to the increased mobility of atoms and radicals on warm grains, producing large molecules *(17)*. It is not clear to what extent these compounds are preserved in the protoplanetary disk. However, chemical models show that complex organic molecules are also efficiently formed on grain surfaces in the disk thanks to the intense UV illumination from the newly born star *(17)*. The presence of interstellar-like organic compounds in comets suggests that they are preserved material synthesized in the outskirts of the solar nebula or at earlier stages of solar system formation.

In Fig. 2, the abundances (normalized to methanol) of nine organic molecules measured in comets Lovejoy and Hale-Bopp are compared to those measured in two interstellar sources. Orion-KL is the best studied hot-core source, because of its proximity and molecular wealth. Quantitative comparison with comets is restricted to methanol, ethanol ($C_2H_5OH$), ethylene glycol [$(CH_2OH)_2$], and glycolaldehyde ($CH_2OHCHO$) measured at the same position in Orion-KL *(18)*. The ethylene glycol abundance in this source is much lower than in the four comets, where it has been detected and this conclusion also applies to the hot-core Sgr B2(N) *(8, 18)*. Glycolaldehyde has not been detected in Orion KL *(18)*, but the derived abundance upper-limit (not shown in the figure) is also much lower than the value measured in comet Lovejoy. IRAS 162293-2422 is the best studied hot corino. The two components of this solar-type binary protostar exhibit large compositional differences, with sources A and B being richer in nitrogen and oxygen-bearing molecules, respectively *(10)*. There is an overall agreement between source B and comet organic composition (Fig. 2), although comets are much richer in ethylene glycol (by 2 orders of magnitude) and ethanol (by a factor 6), whereas they are depleted in ketene ($CH_2CO$). Contrarily to Orion-KL, IRAS 162293-2422(B) *(19)* is moderately underabundant (by a factor 2.5) in glycolaldehyde compared to comet Lovejoy. The same properties are observed for other hot corinos *(20)*.

Comparison with protoplanetary disks is more difficult. The molecules observed in pro-toplanetary disks are mainly small species and associated isotopologues. The most complex identified species are $H_2CO$, $HC_3N$, $CH_3CN$ and $c-C_3H_2$ *(21–24)*. The $HCN/HC_3N/CH_3CN$ abundances are estimated to be in 1/0.05–0.2/0.05–0.2 proportion in the MCW 480 disk, at radial distances corresponding to the comet formation zone *(24)*. This compares well with ratios measured in comets, for example 1/0.02/0.16 and 1/0.08/0.08 in comets Lovejoy and Hale-Bopp, respectively. A comparison with COMs abundances in protoplanetary disks *expected* from chemical models *(25)* is shown in Fig. 2. In addition to grain-surface reactions, reactive desorption and irradiation of ice mantle material in the mid-plane of the disk are included in the chemical model, and enhance the abundances of the more complex molecules in the disk grains *(25)*. Ethylene glycol is, however, produced with a negligible abundance in this model, which implies that efficient formation routes must be found to explain its large abundance in comets *(25)*. Whereas ethylene glycol formation is investigated via the addition of two $CH_2OH$ radicals *(25)*, an alternative grain-surface formation route in protoplanetary disks involves sequential atom-atom additions *(26)* which reaction rates and reaction barriers need to be measured for implementation in chemical models. Altogether, the high COMs abundances in cometary ices, often higher than hot-corino abundances (Fig. 2), are in line with their synthesis through grain-surface reactions and ice irradiation in the early Solar nebula.

Although the possible role of comets in the delivery of terrestrial water has recently been questioned in favor of an asteroidal source *(27)*, the complex organic molecules found in comets, which have their origin in the volatile phase of the solar nebula, are likely related to those present in asteroids *(28)*. The identification of complex organic molecules of biological importance in cometary ices is thus an important step towards our understanding of the origin of life on Earth.

## *Materials and Methods*

### Observations

Comet C/2014 Q2 (Lovejoy) was observed with the IRAM-30m radio telescope during two periods: on 13.8, 15.8 and 16.8 January 2015 (geocentric distance $\Delta = 0.496 - 0.528$ AU) and on 23.7, 24.7, 25.7 and 26.7 January 2015 ($\Delta = 0.624 - 0.675$ AU) with very good weather conditions. The heliocentric distances were 1.31 and 1.29 AU, respectively. Perihelion was on 30.07 January UT, at 1.290 AU from the Sun. The 13–25 and 26 January observations were obtained with the EMIR 1-mm and 3-mm receivers respectively. The main backend we used is a Fourier-transform spectrometer, which covers a frequency range of 2×8 GHz (two side bands separated by 8 GHz, each in two linear polarizations) in a single setup, with a high spectral resolution of 200 kHz. The spectral resolution (0.3–0.2 km s$^{-1}$ when converted into Doppler velocity) allows to resolve the velocity profile of the narrow cometary lines (~ 2 km s$^{-1}$). Using three different tunings we covered the 210–218, 225–233 and 240–272 GHz frequency ranges. Given the excitation conditions encountered in cometary comae at 1–1.5 AU from the Sun, most of the molecules have their strongest lines in this frequency domain. Complex cometary molecules have many lines expected to have similar intensities in this domain and we obtained secure detections when averaging the signal of the lines expected to be the strongest according to our models *(8)*. The number of lines detected with a signal-to-noise ratio larger than three is respectively: 2 for $CH_2OHCHO$, 8 for $C_2H_5OH$, 2 for $HCOOCH_3$, 37 for $CH_3CHO$, 7 for $(CH_2OH)_2$, 9 for HCOOH, 10 for $NH_2CHO$ and 4 for HNCO. Seven lines were used for the upper limit on ketene ($CH_2CO$).

The half power beam width the IRAM-30m at these frequencies ranges from 9.1" to 11.6", which corresponded to 3300 to 5400 km at the comet distance. The pointing accuracy, which was regularly checked on reference pointing sources and also on the comet with coarse maps of the strongest lines, was better than 2".

The time variation of the activity of the comet could be monitored thanks to the presence of several strong CH$_3$OH lines in each setup. We did not see any variation larger than ±20% during the observations (Q$_{CH3OH}$ = 1.2 ± 0.2 × 10$^{28}$ molec. s$^{-1}$, 13–16 January and 1.4 ± 0.2 × 10$^{28}$ molec. s$^{-1}$, 23–25 January). This justifies the averaging of several days of observations to improve the signal-to-noise ratio and derive more precise production rates. Table S1 provides the list of the strongest ethanol and glycolaldehyde lines identified in the spectra and the production rates derived for each line. Several other lines are individually present only at the 1–3 $\sigma$ level but yield a detection when combined altogether and were used to compute production rates.

**Excitation models and production rate determination**

For all molecules we used the latest spectroscopic data available in the JPL *(31)* or CDMS *(32)* database, both for line identification and computation of line strengths. In our model the population distribution of the ground vibrational state of molecule slowly evolves from Local Thermal Equilibrium at the temperature $T_{gas}$ maintained by collisions with neutrals and electrons close to the nucleus *(11,12, and references therein)* to fluorescence equilibrium in less dense parts of the coma. Collisional rates and electron density are based on a water production rate of 5 × 10$^{29}$ molec. s$^{-1}$. We derived $T_{gas}$=73 K from the relative line intensities of the strong methanol lines (Figs. S1 and S2). For the other molecules the line intensity ratios (or rotational temperature) are in agreement with the model (e.g., $T_{rot}$(CH$_3$CHO) = 67 ± 15 K for 68 K predicted). Except for a few molecules like methanol or linear molecules we did not take into account the pumping of the rotational levels by the fluorescence of the infrared vibrational bands, but this process is not expected to play a major role within the ~5000 km of the coma sampled here.

The local gas density is described with the Haser model, which assumes isotropic radial outgassing at a constant velocity. The mean gas expansion velocity of 0.80 km s$^{-1}$ has been derived from the shape of the lines with the highest signal-to-noise ratio: e.g. the strong CH$_3$OH(5$_{+2}$ − 4$_{+1}$E) line at 266.838 GHz, observed both with the FTS and the VESPA autocorrelator (resolution 40 kHz), has a line width of 1.61 ± 0.08 km s$^{-1}$, suggesting a mean velocity of 0.80 ± 0.04 km s$^{-1}$. Given the expansion velocity and beam size, the observations sampled relatively "fresh" molecules, which left the nucleus less than 1.8 h prior to the observation. We took the molecular lifetimes from *(3,6,7,33)*, but at this heliocentric distance (1.3 AU) the derived production rates are not very sensitive to the photodissociation rate of the molecule, provided that its lifetime at 1 AU is larger than 4000 s. The line intensities are computed with our radiative transfer code for a Gaussian beam. Lines are optically thin, implying that the production rates are proportional to the line intensities. When several lines of a molecule are observed, the production rate was computed doing a weighted mean of the production rates computed for each individual line, since not all the lines have the same noise level. All lines which intensity is expected to be ≥ 1$\sigma$ and which are within the observed spectral range are taken into account, even if not detected, to avoid biasing the production rate estimate. Spectra shown in Fig. 1 were obtained by averaging only the stronger lines, weighted according to the noise of each individual spectrum.

For ketene, we obtain a marginal 3 $\sigma$ value of 4.3 ± 1.4 × 10$^{25}$ molec. s$^{-1}$ (0.008% relative to water), but the detection of this molecule needs to be confirmed in a brighter comet. Ethanol is expected to be either in the *trans* or *gauche* states, which are treated separately in our model. *Trans* and *gauche* lines yield similar production rates for both conformers ($Q_{E-trans}$ = 4.6 ± 0.6 × 10$^{26}$ molec.s$^{-1}$, $Q_{E-gauche}$ = 3.0 ± 0.4 × 10$^{26}$ molec. s$^{-1}$). The measured ethanol *trans*/*gauche* ratio of 1.5 ± 0.3 (1.0 ± 0.2 if we take only the 14 lines with S/N ≥ 2 from Table S1) corresponds to a conformeric equilibrium temperature of $51^{+11}_{-6}$ K, slightly lower but close to $T_{gas}$ derived above. Assuming a cooler temperature for ethanol will not change significantly the derived abundance. The intensity and derived total production rates for the strongest lines of glycolaldehyde and ethanol (assuming 50% in each state for ethanol) are provided in Table S1. In the case of the NS radical (Table 1) we have computed the production rate and abundance assuming a release from close to the nucleus.

**Water production rate**

The reference water production rate was primarily derived from observations of the OH radical maser lines at 18-cm carried out with the Nançay radio telescope at the same time. During the 12–17 January period, observations centered on the comet and at 3.0' offset positions (Fig. S3) were used to constrain the quenching of the maser inversion *(34)* yielding an average water production rate $Q_{H2O}$ = 5.0 ± 0.2 × $10^{29}$ molec. s$^{-1}$. OH observations performed between 18 and 25 January were all centered on the nucleus so the water production rates are less constrained for this period. Observations of $H_2O$ were performed with the Odin satellite on 30 January – 3 February (Fig. S3). Analysis of both centered and offset positions up to 3' yields an average production rate $Q_{H2O}$ = 7.5 ± 0.3 × $10^{29}$ molec. s$^{-1}$. Since several methanol lines were detected in each observing setup with the IRAM-30m, we also used the methanol production rate as a proxy for the evolution of water outgassing, correcting possible biases due to larger point- ing uncertainties during the period 23–25 January (no coarse mapping was available). From the interpolation of the Nançay and Odin measurements we estimate that during the period of 23–25 January, $Q_{H2O} \approx 6.0 \times 10^{29}$ molec. s$^{-1}$. The uncertainty on the water outgassing rate should be less than 10%, but we cannot fully exclude that part of the water vapor observed in the 2–8' Odin and Nançay beams comes from the sublimation of icy grains conducting to an overestimate of the water outgassing from the nucleus surface. Nevertheless we do not find clear evidence for a distributed source of gases in the coma of comet Lovejoy: coarse maps of HCN and $CH_3OH$ up to 20" from the nucleus show no evidence for a distributed source, and the interpretation of the Odin water maps does not require to consider a distributed source of $H_2O$ beyond 1'.

## *Supplementary Materials*

**Table S1. Observed strongest lines and production rates.**

**Table S2. Abundances relative to water.**

**Fig. S1. Rotational diagram of the methanol lines observed between 248 and 256 GHz.**

**Fig. S2. Rotational diagram of the methanol lines observed around 242 GHz.**

**Fig. S3. Water production rate: OH and $H_2^{16}O$ observations.**

## *References and Notes*

1. Herbst, E., van Dishoeck, E. F. *Ann. Rev. Astron. Astrophys.* **47**, 427-480 (2009)

2. Bockelée-Morvan, D., Crovisier, J., Mumma, M. J. and Weaver, H. A. In *Comets II*, eds. M.C. Festou, H.U. Keller, and H.A. Weaver (Tucson: University of Arizona Press), 391-423 (2004)

3. Crovisier, J. et al. *Astron. Astrophys.* **418**, 1141-1157 (2004)

4. Elsila, J. E., Glavin, D. P., and J. P. Dworkin. *Meteoritics Plan. Sci.* **44**: 1323-1330 (2009)

5. Materials and methods are available as supplementary materials on *Science* online.

6. Bockelée-Morvan, D. et al. *Astron. Astrophys.*, **353**, 1101-1114 (2000)

7. Crovisier, J. et al. *Astron. Astrophys*. **418**, L35-L38 (2004)

8. Biver, N. et al. *Astron. Astrophys*. **566**, L5 (2014)

9. Belloche, A., Müller, H. S. P., Menten, K. M., Schilke, P. and Comito, C. *Astron. Astrophys*. **559**, A47 (2013)

10. Caux, E. et al. *Astron. Astrophys*. **532**, A23 (2011)

11. Biver, N. et al. *Astron. J.* **120**, 1554-1570 (2000)

12. Biver, N. et al. *Astron. Astrophys*. **449**, 1255-1270 (2006)

13. Pearson, J. C., Brauer, C. S., Drouin, B. J. *J. Mol. Spec*. **251**, 394-409 (2008)

14. Paganini, L. et al. *Astrophys. J.* **791**, 122 (2014)

15. Vastel, C., Ceccarelli, C., Lefloch, B. & Bachiller, R. *Astrophys. J.* **795**, L2 (2015)

16. Charnley, S. B. *Mon. Not. Roy. Astron. Soc.* **291**, 455 (1997)

17. Garrod, R. T., Weaver, S. L. W. and Herbst, E. *Astrophys. J.* **682**, 283-302 (2008)

18. Brouillet, N. et al. *Astron. Astrophys*. **576**, A129 (2015)

19. Jørgensen, J. K. et al. *Astrophys. J.* **757**, L4 (2012)

20. Taquet, V. et al. *Astrophys. J.* **804**, 81 (2015)

21. Dutrey, A., Guilloteau, S. and Guelin, M. *Astron. Astrophys*. **317**, L55-L58 (1997)

22. Chapillon, E. et al. *Astrophys. J.* **756**, 58 (2012)

23. Qi, C., Öberg, K. I., Wilner, D. J. and Rosenfeld, K. A. *Astrophys. J.* **765**, L14 (2013)

24. Öberg, K. I. et al. *Nature* **520**, 198-201 (2015)

25. Walsh, C. et al. *Astron. Astrophys*. **563**, A33 (2014)

26. Charnley, S. B. 1997. In *Astronomical and Biochemical Origins and the Search for Life in the Universe, IAU Colloquium* **161**, C. B. Cosmovici, S. Bowyer, D. Werthimer Eds. (Bologna, 1997) pp. 89–96.

27. Alexander, C. M. O.'D. et al. *Science* **337**, 721 (2012)

28. Alexander, C. M. O.'D. In *The Molecular Universe, Proc. IAU Symposium* **280**, 288-301 (2011)

29. Bisschop, S. E., Jørgensen, J. K., Bourke, T. L., Bottinelli, S. and van Dishoeck, E. F. *Astron. Astrophys*. **488**, 959-968 (2008)

30. Fray, N., Bénilan, Y., Cottin, H. and Gazeau, M.-C. *J. Geophys. Res*. **109-E7**, E07S12 (2004)

31. Pickett, H. M. et al. *J. Quant. Spectrosc. & Rad. Transfer* **60**, 883-890 (1998) (http://spec.jpl.nasa.gov/)

32. Müller, H. S. P., Schlöder, F., Stutzki J. & Winnewisser, G. *J. Mol. Struct*. **742**, 215-227 (2005) (http://www.astro.uni-koeln.de/cdms/)

33. Crovisier, J. *J. Geophys. Res*. **99-E2**, 3777-3781 (1994)

34. Gérard, E. et al. *Plan. Space Sci*. **46**, 569-577 (1998)

## *Acknowledgments*

The observations were conducted under the target of opportunity proposal D04-14 and regular proposal 128-14 and we gratefully acknowledge the support from the IRAM director for awarding us discretionary time and the IRAM staff for its support and for scheduling the observations on short notice.

**Funding:** IRAM is supported by INSU/CNRS (France), MPG (Germany) and IGN (Spain).

**Author contributions:** N.B. coordinated the observation program. N.B. and R.M. conducted the IRAM observations, J.C. and A.S. were responsible for the Nançay and Odin comet observations, respectively. P.C. reduced the Nançay OH observations. N.B. carried out the data reduction and modeling of the IRAM and Odin data. D.B.-M. provided the comparative study on complex organics in comet and protostars and other galactic sources of interest. All authors contributed to the observation planning and commented on the manuscript.

**Competing interests:** The authors declare that they have no competing interests.

**Data and materials availability:** The spectral dataset is available at the CDS via anonymous ftp to http://cdsarc.u-strasbg.fr (ftp://130.79.128.5).

## *Figures and Tables*

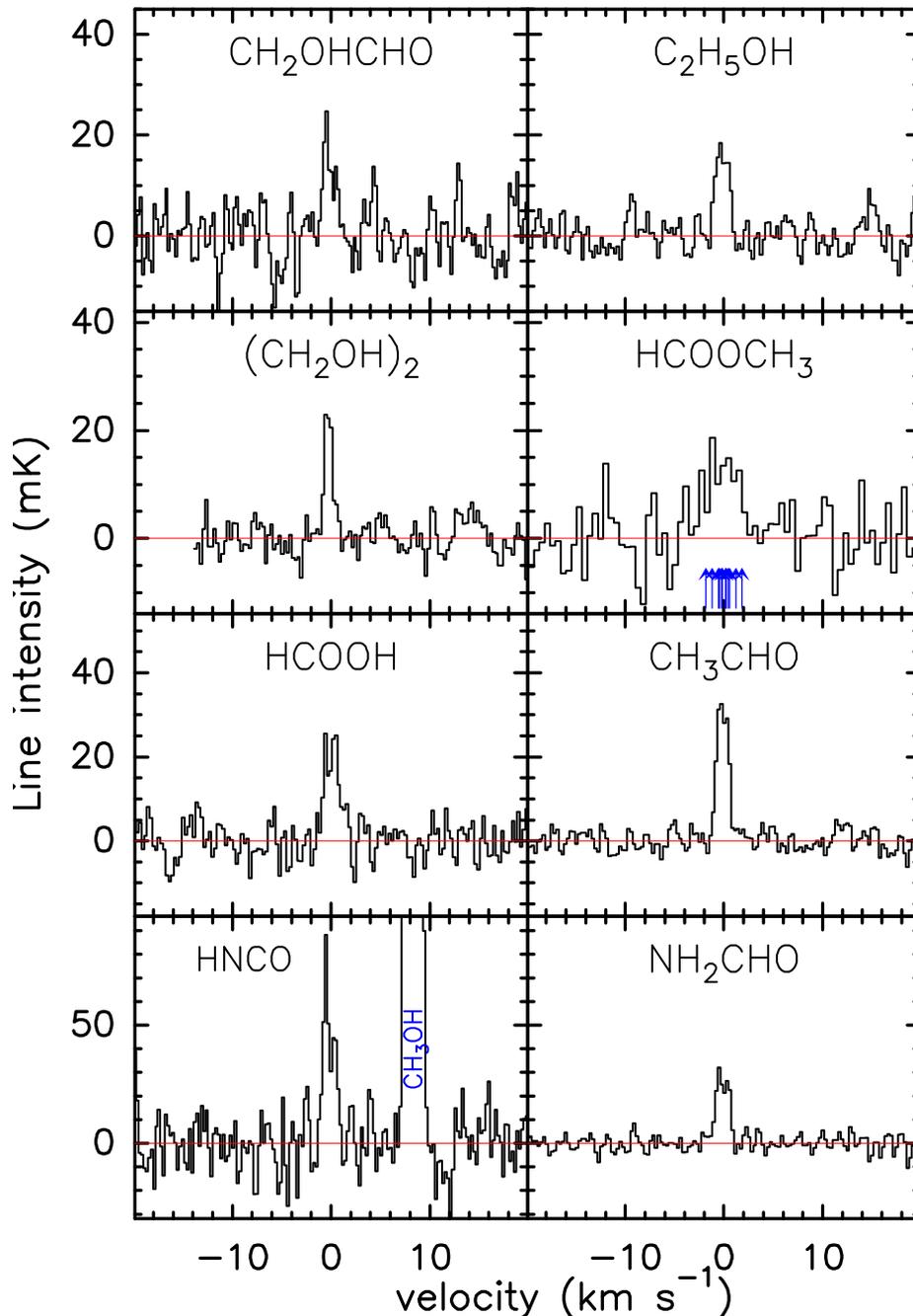

**Fig. 1. Spectra of organics in comet C/2014 Q2 (Lovejoy).** The observations were obtained with the IRAM-30m radio telescope in the 211-272 GHz band between 13 and 25 January 2015. The velocity scale is in the nucleus rest frame. Intensity is given in the main beam temperature scale. Spectra, from top left are: glycolaldehyde (CH$_2$OHCHO, average of two lines), ethyl alcohol (ethanol, C$_2$H$_5$OH, average of 13 lines), aGg' ethylene glycol ((CH$_2$OH)$_2$, average of 14 lines), methyl formate (HCOOCH$_3$, average of two groups of blends of several lines, which positions are marked by blue arrows), formic acid (HCOOH, average of 6 lines), acetaldehyde (CH$_3$CHO, average of 40 lines), isocyanic acid (HNCO($11_{0,11}-10_{0,10}$) line at 241.774 GHz) and formamide (NH$_2$CHO, average of 10 lines). The signal-to-noise ratio is 6 for glycolaldehyde, 10 for ethanol and larger than 7 for the other molecules.

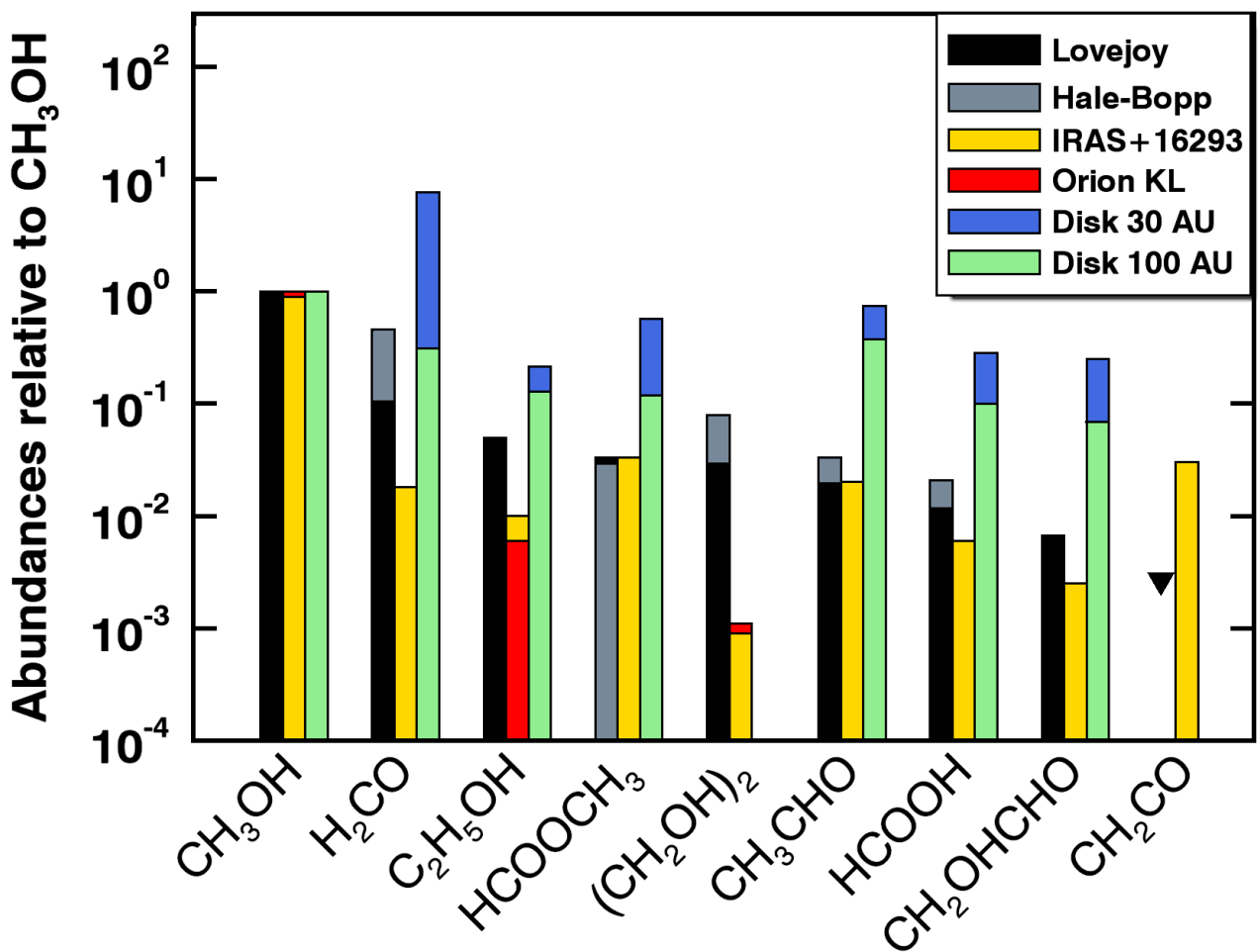

**Fig. 2. Abundances of complex organics in comets and protostars.** The abundances measured in comets Lovejoy (Table 1) and Hale-Bopp (Table S2) are compared with those measured in the low-mass protostar IRAS 16293-2422(B) *(19, 29)* and in the high-mass protostar Orion-KL *(18)*, and with results from chemical modeling in a proto-planetary disk *(25)*. Abundances are given relative to $CH_3OH$, the most abundant organic molecule in these sources. $H_2CO$ values may be less relevant, since, for comets, a significant fraction is released from grains *(30)*, whereas for IRAS 16293-2422, the plotted value pertains to the sum of the contributions to the two components A and B of the binary source. For $CH_2CO$, the black triangle is a 3 $\sigma$ upper limit obtained in comet Lovejoy.

Table 1. Abundance of molecules detected in comet Lovejoy.

| Abundances relative to water (%) | | | | | |
|---|---|---|---|---|---|
| CHO molecules | | Nitrogenous molecules | | Sulphureted molecules | |
| CO | 1.8 | HCN | 0.09 | $H_2S$ | 0.5 |
| $H_2CO$ | 0.3 | HNC | 0.004 | OCS | 0.034 |
| $CH_3OH$ | 2.4 | HNCO | 0.009 | $H_2CS$ | 0.013 |
| HCOOH | 0.028 | $CH_3CN$ | 0.015 | CS | 0.043 |
| $(CH_2OH)_2$ | 0.07 | $HC_3N$ | 0.002 | SO | 0.038 |
| $HCOOCH_3$ | 0.08 | $NH_2CHO$ | 0.008 | NS | 0.006 |
| $CH_3CHO$ | 0.047 | | | | |
| $CH_2OHCHO$ | 0.016 | | | | |
| $C_2H_5OH$ | 0.12 | | | | |

Abundances relative to water of the 21 molecules detected in comet Lovejoy are based on the average of production rates ratios for the 13-16 and 23-25 January periods. The uncertainty on the abundances, taking into account all sources of errors, including uncertainty on the water production rate, is below 20% except for NS for which it is 30% (4 $\sigma$ detection).